# SIMULATION OF THE GAMMA-RAY GALACTIC DISTRIBUTION AS SEEN BY THE AMS-02

**M. Mollá[1], J. Alcaraz[1], J. Berdugo[1], J. Bolmont[2], J. Casaus[1], E. Lanciotti[1], C. Mañá[1], C. Palomares[1], E. Sánchez[1], F.J. Rodríguez[1], M. Sapinski[2], I. Sevilla[1], A.S. Torrentó[1], on behalf of the AMS collaboration**

[1] C.I.E.M.A.T., Ministerio de Educación y Ciencia, Avda. Complutense 22, 28040-Madrid, Spain

[2] Groupe d'Astroparticules de Montpellier, Université de Montpellier II, Montpellier, France

## ABSTRACT

In this note, we report the results of a simulation of the galactic component of high energy gamma rays, as seen by the future AMS-02 experiment on-board the International Space Station. The purpose of AMS is to measure the Cosmic Ray spectra of numerous particles with unprecedented accuracy.

This simulation has been performed with the GALPROP code and its output interfaced with the AMS Fast Simulator, to produce a sky-map of yearly counts in the AMS energy range (over 1 GeV).

## 1. INTRODUCTION

The Alpha Magnetic Spectrometer (AMS) (Aguilar et al. 2004) is a particle physics experiment designed for space operation. It will collect a huge amount of statistics of primary cosmic rays with unprecedented sensitivity from the International Space Station. In order to achieve its goal, AMS is composed of five sub-detectors, which provide complementary information about the characteristics of the incoming particles. These sub-detectors include the following: a Transition Radiation Detector (TRD); a Silicon Tracker Detector (STD) operating inside a superconducting magnet; a Ring Imaging Cherenkov detector (RICH); a Time-Of-Flight (TOF) system and an Electromagnetic Calorimeter (ECAL). Many of them involve important developments in the technologies used, the magnet being the first of its kind in space.

A prototype (AMS-01) was tested on-board the STS-91 flight of the Space Shuttle. During 10 days, this simplified version of the experiment performed successfully and obtained important science results, like the two-component spectrum of low-orbit cosmic rays and the new upper limit for the anti-helium/helium ratio (Aguilar et al. 2002). AMS-02 will be placed on the International Space Station (ISS) for a minimum three-year operation period, starting in 2007.

## 2. COSMIC RAYS IN THE GALAXY

Cosmic Rays (CR) are high-energy particles of extraterrestrial origin. Charged CR are mainly constituted by an isotropic flux of protons, helium and electrons. Gamma rays and neutrinos are also part of the cosmic radiation keeping directional information of their sources due to their neutral nature. The origin of charged CR up to ~100 TeV is believed to be supernovae explosions.

The diffuse galactic flux of gamma rays may come from three different origins:

a) the interaction of primary CR with the interstellar medium (producing pions which decay into gamma rays);
b) the up-scattering of the interstellar photon field by accelerated electrons (inverse Compton process);
c) the braking radiation from electrons in the electrostatic field of the interstellar gas (bremmstrahlung).

For this study the GALPROP code (Strong & Moskalenko 1998, Strong et a. 2003) was used in order to simulate the CR propagation and interactions, according to the author's latest model (Strong et al. (2004)). A recent interstellar diffuse gas radial distribution was used (Nakanishi & Sofue 2004). The gas distribution is a key component for determining the flux of gammas from pion decay and bremmstrahlung. The new distribution, as we can see in Fig.1, is clearly different from previous estimates. The effect on final results of GALPROP is important, see also Mollá et al. (2004).

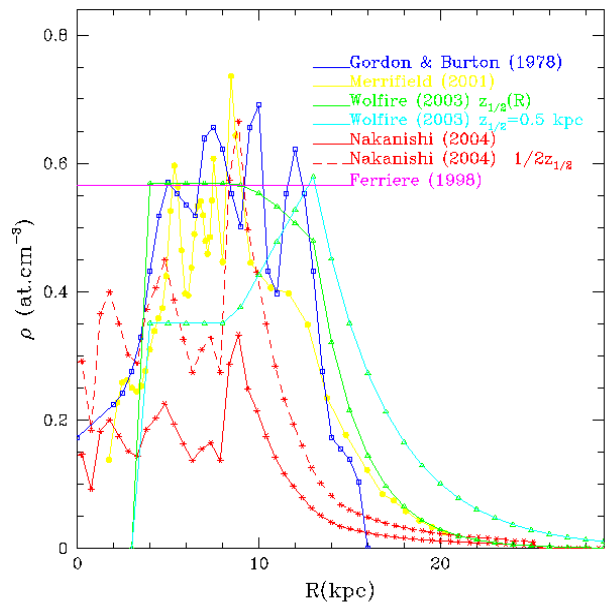

Fig1. Interstellar diffuse gas density radial distribution as obtained from Nakanishi & Sofue (2004) compared with other author's data as labelled in the figure.

## 3. GAMMA RAY DETECTION WITH AMS

Charged CR are the primary goal of the AMS experiment. However, it is also possible to use two of their main sub-detectors for high-energy photon detection by pair conversion (Battiston et al. 2000, Lamanna 2002, Aguayo et al. 2003). A scheme of the detector layout is shown in Fig. 2.

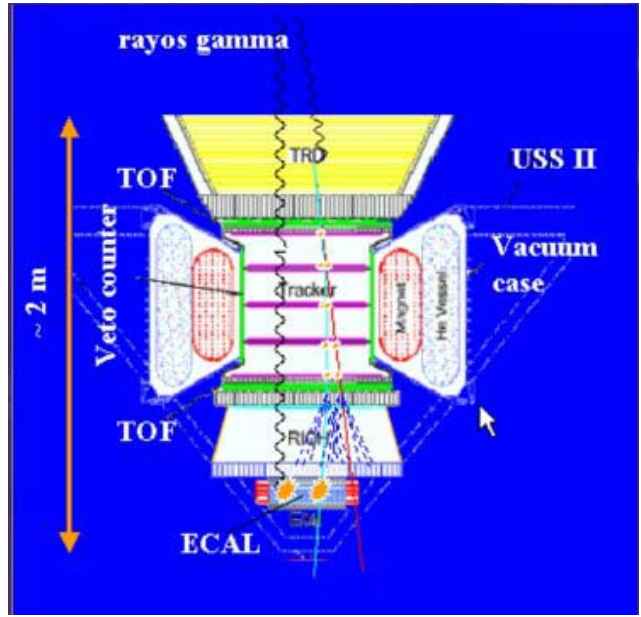

Fig.2. Gamma ray detection in AMS-02 and detector layout. The two methods of photon detection are schematically shown.

An energetic photon has a high chance of converting into a pair as it traverses the TRD (which amounts to about 0.3 radiation lengths). The STD ("tracker" in Fig.2) provides a very precise measurement of the gamma-ray direction by detecting directly the converted electron-positron pair. Their path through the detector is reconstructed by means of the energy deposited in each of the silicon layers, from which we can derive the original photon's direction and its momentum (the particles' tracks bend inside the magnetic field).

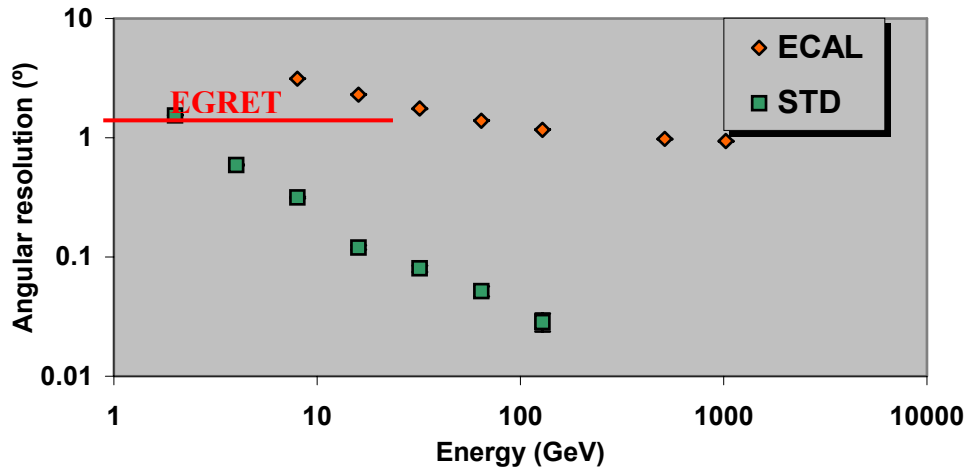

Fig.3 AMS single photon angular resolution from MonteCarlo simulation, defined as the angular width from the true direction where 68% of the photons are contained. The horizontal line represents EGRET´s approximate performance.

With this method, we can reach a precision of around 0.02 degrees for single photon angular resolution (Fig.3), thanks to the fine precision of the STD strips, of the order of tens of microns.

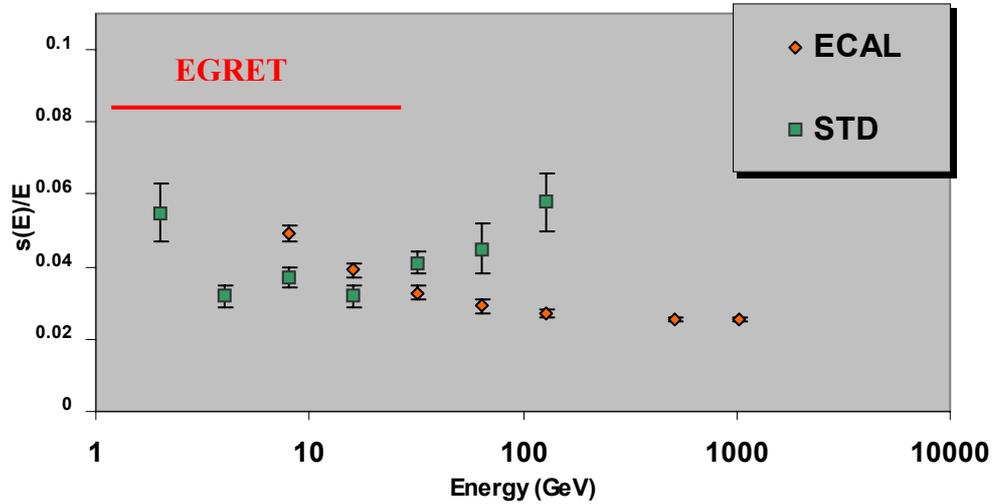

Fig.4. The energy resolution as relative error, for both AMS sub-detectors as compared with the approximate performance for EGRET.

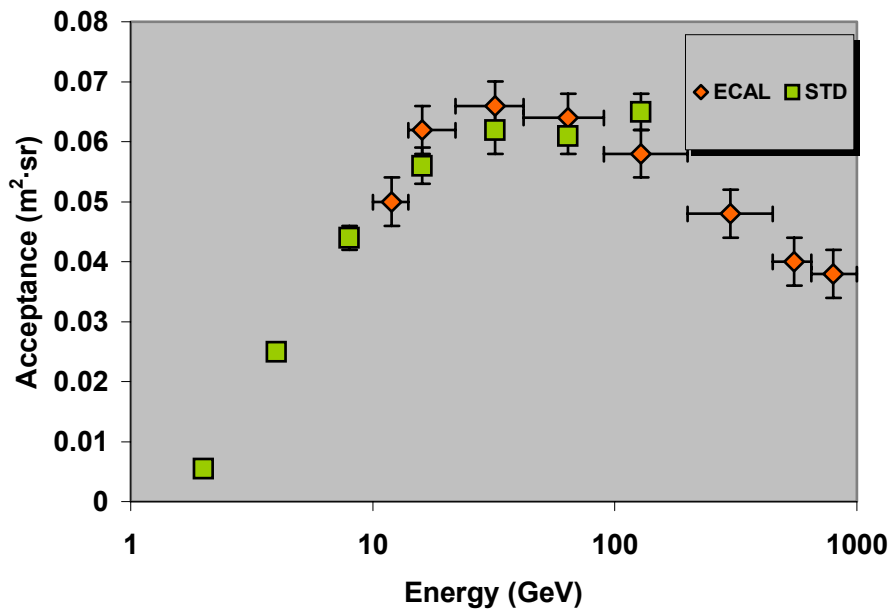

Fig.5. Gamma-ray acceptance for both sub-detectors, ECAL and STD

An unconverted gamma photon can be nonetheless detected by the ECAL. This detector is composed of several super-layers of lead and scintillating fibres connected to photo-multipliers forming a fine matrix. Thus, the incoming gamma energy can be determined with ~5% error, (see the energy resolution in Fig. 4) by analysing the electromagnetic shower development of the gamma ray.

Another fundamental figure of merit of any gamma-ray observatory is its acceptance in area*solid angle units. The ECAL energy range is limited at low energies by high backgrounds and at high energy by detector saturation. STD, on the other hand, is limited by the loss of one of the pair's components at low energy (spiralization of the path in the magnetic field) and the intrinsic tracker resolution at the high end of the energy spectrum.

Fig.5 shows the results from a MonteCarlo simulation of the global acceptance for both ECAL and STD. Auxiliary systems such as a twin CCD-star-tracker and a GPS system will complement these capabilities.

## 4. SIMULATION RESULTS FOR THE GALACTIC PROPAGATION MODEL

The gamma ray model from GALPROP was interfaced with a simulation specifically designed to evaluate the gamma-ray detection capabilities of AMS (using some functions from the AstroROOT software package from the INTEGRAL team, see R.Rohlfs´s AstroROOT webpage).

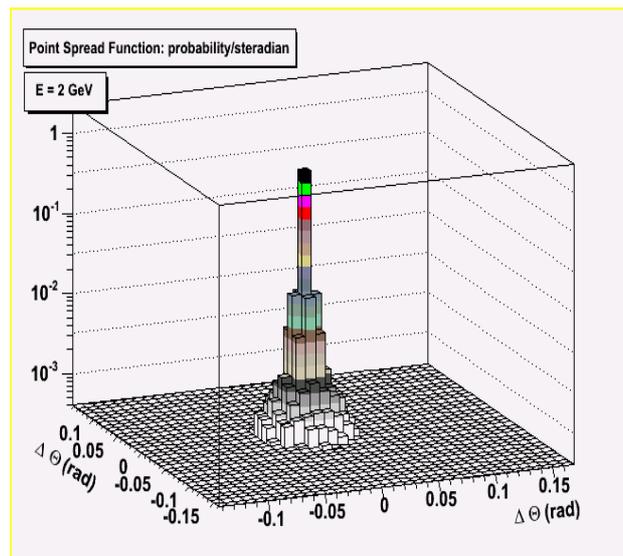

Fig.6. Point Spread Function for 2 GeV.

The GALPROP code is able to provide a FITS galactic map at certain energy of the diffuse galactic gamma ray fluxes. These maps are available for each of the processes producing gamma photons from interactions of cosmic rays with the galactic medium (see section 2). The flux at each coordinate is convoluted with the acceptance of the detector for that energy (Fig. 5), the exposure time (determined by an orbit simulation), and the angular point spread function (Fig. 6). Thus, by adding up the maps for all processes and energies, we obtain the simulated inner Galaxy count map for γ-rays after a three-year exposure period, such as we show in Fig.7.

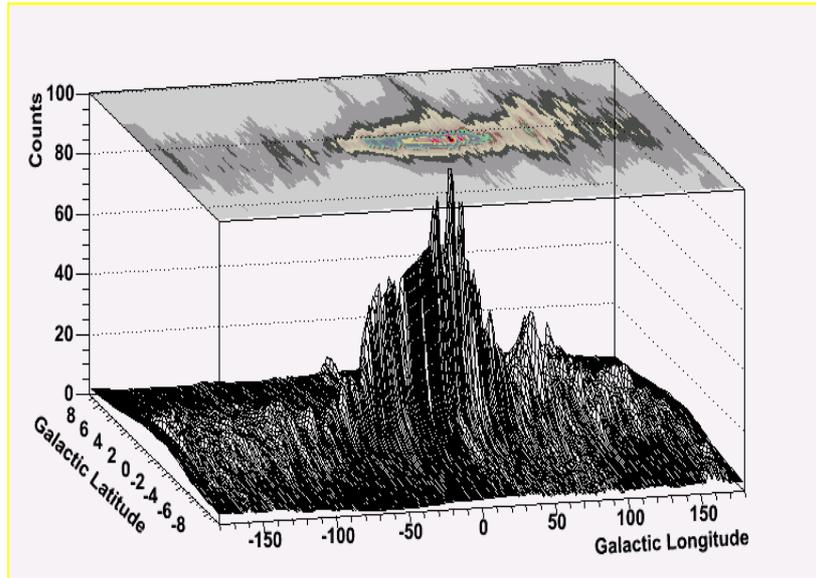

Fig.7. Simulated inner Galaxy count map (energy > 1GeV) for γ-rays after a three-year exposure period. The coordinate binning is 0.5 degrees.

From this result, we can derive the predicted gamma-ray spectrum and the statistical errors (Fig. 8). The solid line represents the theoretical model provided by the GALPROP simulation, for two Galactic regions. Superimposed are the statistical errors, which will be present in the estimations of AMS. The solid points with error bars are the data from EGRET. The most important contribution of AMS to the Galactic spectrum will be extending the energy range beyond of ~20 GeV, the EGRET cap, especially at the outer Galaxy regions.

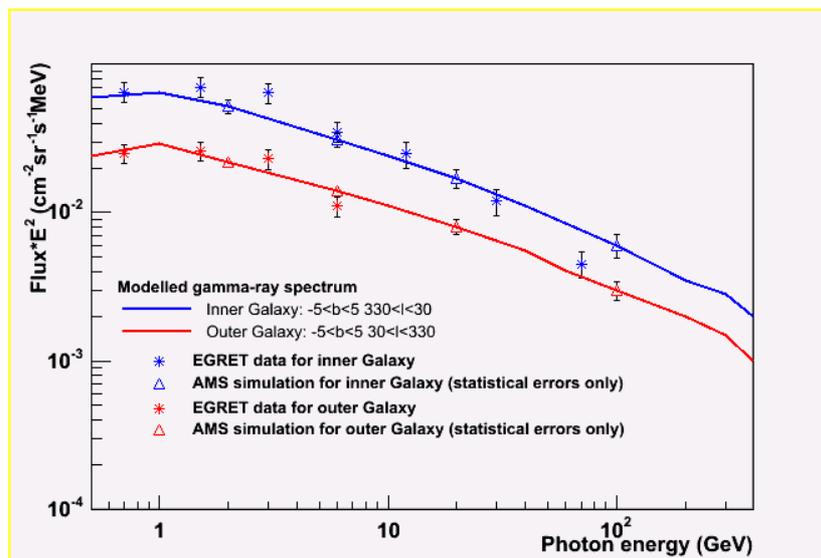

Fig.8 Gamma ray spectrum predictions and errors provided by the AMS gamma fast simulator (triangles with error bars) and EGRET data (asterisks).

We have also obtained the intensity profiles for some energy ranges in galactic longitude and latitude. The intensity profile in longitude for energies over 10 GeV is shown in Fig. 9. Here it becomes evident that pion production (and therefore the gas distribution) is the main contributor to diffuse Galactic emission, though according to Strong et al. (2004) inverse Compton can become very important at higher energies (~100 GeV). AMS will explore this energy range as well.

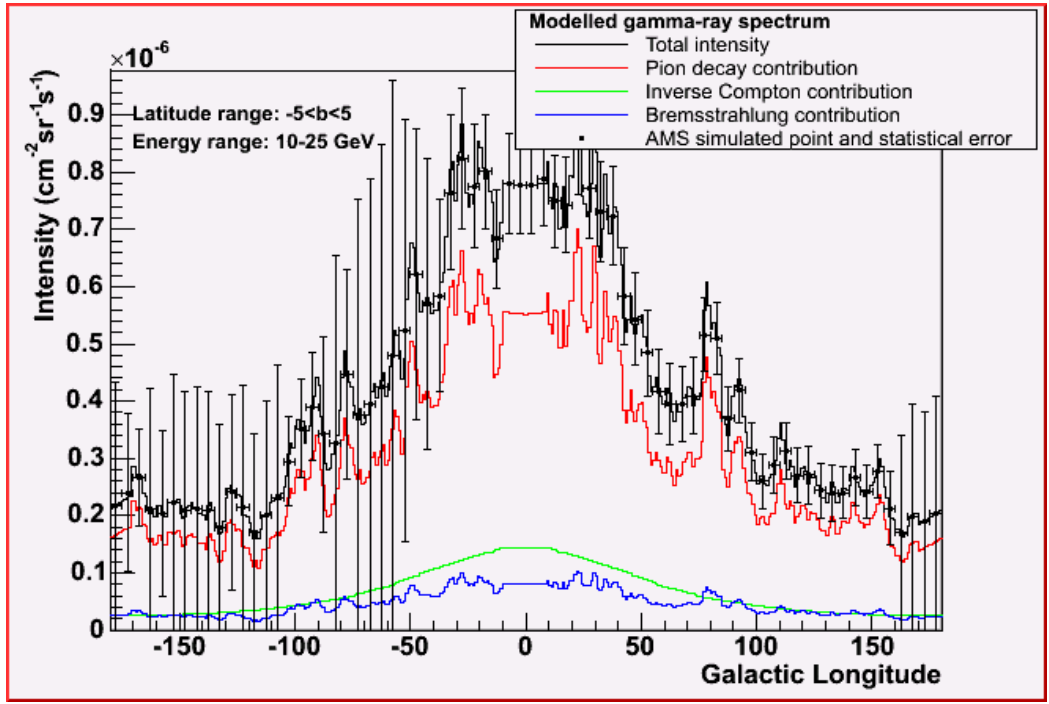

Fig. 9. Galaxy intensity profile for energy over 10 GeV.

Other important studies such as sensitivity to galactic and extragalactic gamma sources and dark matter gamma-ray signatures will be reported in the near future. In this context, the study of the galactic background contribution is of utmost importance to evaluate the significance of a weak signal.

**SUMMARY**

The aims for this study were to analyse the possible contribution of AMS to the study of the diffuse galactic gamma ray distribution and to evaluate the contribution of the gamma ray background to other types of signals.

We conclude that the AMS, a charged cosmic ray detector, will significantly contribute in the field of gamma ray astronomy, specially considering its high angular resolution (~0.02 degrees), comparable to other specialized observatories such as GLAST at high energies, though with its smaller acceptance. Energy resolution also proves to be an important asset with the ECAL detector, reaching a 5%.

We show that it will be possible to extend the spectrum to higher (>50 GeV) energies, an interesting range to study the underlying processes of gamma ray production and their relationship with different cosmic ray species;

BIBLIOGRAPHY

Aguayo, P. et al.(2003) Proc. XXIX Bienal RSEF, 369
Aguilar, M. et al. [AMS 02 Collaboration] (2004) submitted to NIM A.
Aguilar, M. et al. [AMS 01 Collaboration] (2002) Phys.Rep. 366, 331.
Battiston, R. et al. Astropart.Phys. (2000) 13, 51
Dickey, J. M. & Lockman, F.J. (1990) A.R.A.A., 28, 215
Ferrièrre, K. (1998) ApJ., 497, 759
Gordon, M.A. & Burton, W.B. (1976) ApJ., 208, 346
Jones, F.C., Lukasiak, A., Ptuskin, V., & Webber, W. (2001) ApJ, 547, 264
Lamanna, G. (2002) Nucl. Phys.B (Proc.Suppl.)113,177
Maurin, D., Donato, F., Taillet, R. & Salati, P. (2001) ApJ., 555, 585
Mollá, M., Aguilar, M., Alcaraz, J., et al. (2004) ESP Astrophysics Symposia Series, Springer-Verlag, in press
Nakanishi, H. & Sofue, Y. (2004) PASJ, 55, 191
Olling, R.P. & Merrifield, M.R. (2001) MNRAS, 326, 164
Rohlfs,R. AstroROOT webpage http://isdc.unige.ch/Soft/AstroRoot/
Strong, A.W. & Moskalenko, I.V. (1998) ApJ, 509, 212
Strong, A.W. & Moskalenko, I.V. (2003) $28^{th}$ IRC Conf., 1921
Strong, A.W., Moskalenko, I.V. & Reimer, O. (2003) 28 IRC Conf., 1921
Yanasak, N.E., Wiedenbeck, M. E., Mewaldt, R.A. et al. (2001) ApJ., 563, 768
Wolfire, M.G., McKee, C.F., Hollenbach, D., et al. (2003) ApJ, 587, 27